\documentstyle[12pt,epsf]{article}
\def\be{\begin{equation}}
\def\ee{\end{equation}}
\def\rg{renormalisation\ group}

\def\no#1{:\!#1\!:}
\def\ffrac#1#2{\textstyle{#1\over#2}\displaystyle}
\gdef\journal#1, #2, #3, 1#4#5#6{{\sl #1~}{\bf #2}, #3, 1#4#5#6}

\begin{document}
\pagestyle{empty}
\vspace{-1mm}
\begin{flushright}
NSF-ITP-97-056\\
ANU-MRR-015-97\\
cond-mat/9705238
\end{flushright}
\vspace{5mm}
\begin{center}
{\bf \LARGE Extraordinary transition in the\\
 two-dimensional $O(n)$ model\\}
\vspace{8mm}
{\bf \large Murray T. Batchelor\\}
\vspace{2mm}
{Department of Mathematics\\
School of Mathematical Sciences\\
Australian National University\\Canberra ACT 0200, Australia\\}
\vspace{5mm}
{\bf \large John Cardy\\}
\vspace{2mm}
{Department of Physics, Theoretical Physics\\1 Keble Road\\Oxford OX1 3NP, UK\\
\& All Souls College, Oxford}
\end{center}
\vspace{6mm}
\begin{abstract}
The extraordinary transition which occurs in the two-dimensional $O(n)$
model for $n<1$ at sufficiently enhanced surface couplings is studied by
conformal perturbation theory about infinite coupling and by finite-size
scaling of the spectrum of the transfer matrix of a simple lattice
model. Unlike the case of $n\geq1$ in higher dimensions, the surface
critical behaviour differs from that occurring when fixed boundary
conditions are imposed. In fact, all the surface scaling dimensions are
equal to those already found for the ordinary transition, with, however,
an interesting reshuffling of the corresponding eigenvalues between
different sectors of the transfer matrix.
\end{abstract}
\newpage
\pagestyle{plain}
\setcounter{page}{1}
\setcounter{equation}{0}
\section{Introduction}
The behaviour of linear polymers interacting with a surface has been
studied extensively \cite{DL93,E93,C96}. 
In the dilute limit they are modelled on the
lattice by self-avoiding walks, which are themselves related to the
$n\to0$ limit of an $O(n)$ spin model. In two dimensions, in
particular, it has been possible, using methods of conformal field
theory and integrable models, to obtain much exact information
concerning the critical behaviour.

An explicit example is provided by the loop model on the honeycomb
lattice, defined as follows. The partition function is a sum over all
configurations of closed oriented loops on the lattice. A given site of
coordination number 3 is therefore unoccupied (with weight $x$) or has
just one oriented bond entering and leaving the site. There are 6 such
configurations for a site, each occurring with weight 1. The
parameter $x$ is thus the fugacity for empty sites, which controls the
overall monomer density in the polymer language. There is also a
fugacity $y$ for empty sites on the boundary, which have coordination
number 2. In addition, each oriented closed loop carries a factor $n/2$
(so that on summing over both orientations the factor is $n$). This
nonlocal weight is equivalent to inserting local weights $e^{\pm i\chi}$
at each occupied site, with the sign chosen according to whether the walk
turns left or right as it passes through the site, as long as
$n=2\cos6\chi$. These oriented loop configurations then correspond to
the diagrams of the high temperature expansion in powers of $1/x$
of a complex $O(n/2)$ ($\equiv$ real $O(n)$) model, with complex spins
$S_a$ ($a=1,\ldots,n/2$) at each lattice site.

\begin{figure}
\centerline{
\epsfxsize=3.5in
\epsfbox{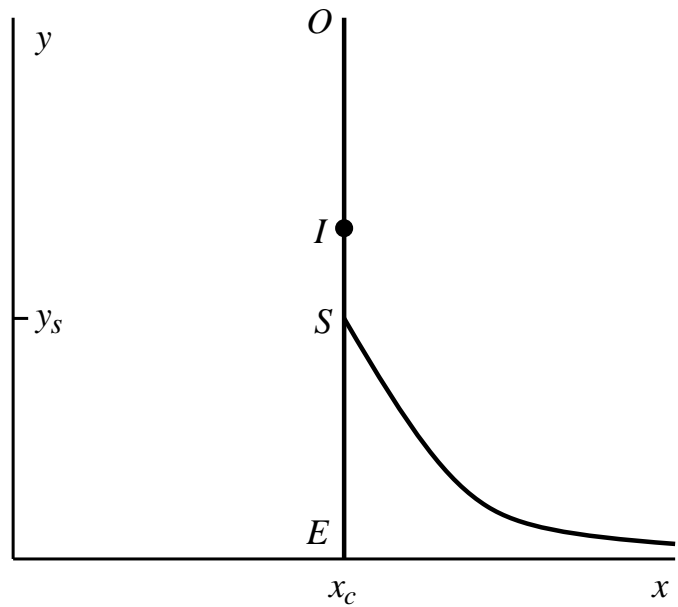}}
\caption{Schematic phase diagram of the $O(n)$ model for $n<1$, with
bulk and surface vacancy fugacities $x$ and $y$ respectively. 
$S$ is the special
transition, from which emerge the lines of ordinary $(SO)$,
extraordinary $(SE)$, and surface transitions. The particular model
considered is integrable at the points $S$, $I$, and $E$.}
\label{pd}
\end{figure}

The schematic phase diagram of this model is shown in Fig.~\ref{pd}.
Only the details depend on the specific model.
There is a bulk critical point at $x=x_c$, where \cite{N82,B86}
\be
x_c = \sqrt{2 + \sqrt{2-n}},
\ee
which controls the
behaviour of long walks of fixed length $N$ in the bulk. This is because
the relation between fixed fugacity and fixed length ensembles is
through a discrete Laplace transform: for example the generating
function $\sum_Nc_Nx^{N}$ for the number of $N$-step walks per site is
given by the susceptibility of the $O(n)$ model, which diverges as $x\to
x_c$ from above. When $y>y_s$, quantities near the surface become
singular at $x=x_c$, but with exponents which differ from those in the
bulk. This is the line $SO$ of ordinary transitions. It terminates at
the point $S$ where $y=y_s$, the special transition. When $y<y_s$
surface quantities actually become singular at a value of $x=x'_c(y)>x_c$, while
the bulk remains nonsingular. This is termed the surface transition, and
is in the universality class of the $(d-1)$-dimensional $O(n)$ model.
However, for $y<y_s$, surface quantities then undergo a second
transition on the line $x=x_c$, since they are coupled to the bulk
critical behaviour. This is the extraordinary transition. The phase
diagram in Fig.~\ref{pd} is believed to be generic to short-range $O(n)$
models, both in three dimensions, and in two dimensions when $n<1$. (For
$n\geq1$ in $d=2$ the surface and ordinary transitions disappear since
the surface cannot order independently of the bulk, being
one-dimensional.) This picture is confirmed by exact results on the
specfic model described above, which turns out to be solvable by Bethe
ansatz methods at $x=x_c$ and for two values of $y$: $y=x_c$ \cite{BS93,YB95a,YB95b} 
which yields values for the surface exponents in agreement with those expected
for the ordinary transition on the basis of conformal field theory
arguments \cite{C84a,DS86}; and $y=y_s$, where \cite{BY95a,YB95a,YB95b}
\be
y_s = { \sqrt{2-n} \over \sqrt{2 + \sqrt{2-n}}},
\ee
at which the exponents are different. This latter
point is therefore identified with the special transition.

Note that the extraordinary transition is irrelevant for the fixed $N$
ensemble of dilute polymers, since its large $N$ behaviour is controlled
by the largest singularity in $x$, which in this case is the surface
transition. Physically this is because for $y<y_s$ a finite length
polymer will bind to the surface and therefore be governed by
$(d-1)$-dimensional exponents. Conversely, for $y>y_s$ it will explore
the bulk and have an effective repulsion from the surface. The special
transition at $y=y_s$ is therefore important in that it describes the
absorption transition for a large but finite length polymer at the
boundary.

Nevertheless, within the fixed fugacity ensemble, which is certainly
realisable at least in simulations and transfer matrix calculations, the
extraordinary transition is accessible, and one may enquire as to its
universal properties. For conventional spin models, including the $O(n)$
model with $n\geq1$ in $d>2$, it may be argued that the extraordinary
transition is identical to that which occurs in the presence of a surface
ordering field. This is because for temperatures below that of the
surface transition the surface is expected to be ordered, which means
that the $O(n)$ symmetry is spontaneously broken. However, once it is
broken, it should not matter whether this breaking was the result of an
explicit surface symmetry-breaking field. For Ising-like
systems, there is ample evidence that these two problems are indeed in
the same universality class. The limit of infinite surface ordering
field is equivalent to the case of fixed boundary conditions. 

An analogous line of argument has led the authors of Refs.~\cite{BEG89,BE94}
to analyse the extraordinary transition of the $O(n)$ model in the limit
$n\to0$ in terms of the same model with \em fixed \em boundary
conditions. However, this is not correct, because in the absence of
explicit symmetry breaking fields the $O(n)$ model cannot exhibit
surface order for $n<1$, even when $x<x_c'$. For, if the spins near the
boundary were to be aligned in some fixed direction, walks in the
vicinity of the boundary would be counted with weight 1, rather than
$n$. The situation is very analogous to the case when $x<x_c$ in the
bulk: in that case, the $O(n)$ spins do not order and the symmetry is
not broken. Rather this corresponds to the so-called dense phase, when a
single polymer occupies a finite fraction of the sites in a finite
volume. Similarly, $x<x_c'$ corresponds to the `dense' phase on the
surface (which, being one-dimensional, is rather trivial.) There is,
nevertheless, an extraordinary
transition in this dense phase when the bulk goes critical.

\begin{figure}
\centerline{
\epsfxsize=2in
\epsfbox{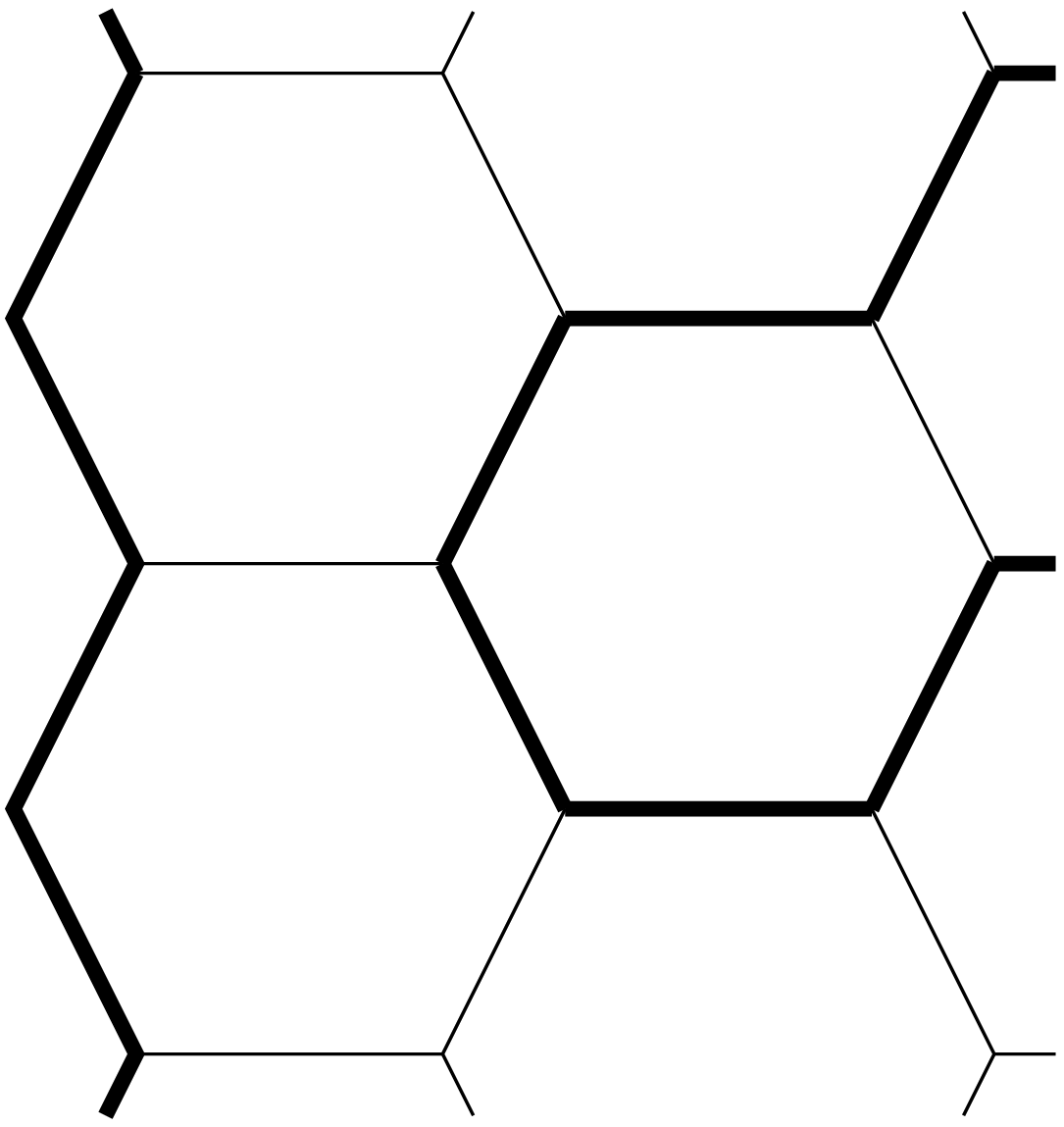}}
\caption{Part of the boundary of the honeycomb lattice. When $y=0$, a
polymer is rigidly bound to the boundary. This means that a second
polymer is effectively repelled from the second layer of sites which
have fugacity $x_c$.}
\label{tef}
\end{figure}

The critical properties of this transition are in fact identical to
those of the ordinary transition. To see this, consider the special
point $y=0$ at the end of the line. Then there are no vacant sites on
the boundary, and if there is just one polymer in the system, it will be
rigidly bound to the surface. However this will mean that other polymers
are in fact repelled from the surface. In the explicit model introduced
above, it may be seen that sites one lattice constant into the bulk from
the boundary will have an effective surface fugacity precisely $y=x_c$,
which is the integrable ordinary point! This is illustrated in
Fig.~\ref{tef}.
For this reason, one may refer
to the point $y=0$ as the `teflon' point: one polymer coats the boundary and
makes it effectively repulsive to other polymers.

The properties at $y=0$ are therefore very simple: they are those of a
lattice with one layer of sites removed, at the ordinary point $y=x_c$.
This leaves the remainder of the line $0<y<y_s$ to be discussed,
and this forms the bulk of this paper. We shall argue that all the
universal properties along this line are the same as those at $y=0$, by
using two sets of arguments. In Sec.~\ref{sec2} we use perturbation
theory in $y$, coupled with various arguments from conformal field
theory, to show that the variable $y$ is irrelevant in the sense of the
\rg, so that all the universal properties are the same, at least for
sufficiently small $y$. In order to perform perturbation theory about
such a critical point it is necessary to impose an infrared cutoff, and
this is most easily realised by considering the system in an infinitely
long strip of width $L$. The universal properties are then encoded in
the finite-size scaling behaviour of the spectrum of the
transfer matrix of this model. This may be studied within
conformal perturbation theory about the teflon point. We find indeed
that the scaling part of the spectrum is unaffected by the perturbation.
However, there are additional degeneracies at $y=0$ compared to those
along the line of the ordinary transition. Some of these are 
resolved by the perturbation, but not in the scaling
part of the spectrum, where the degeneracies persist.
We show how this comes about and derive the
qualitative nature of the splitting.

This analytic work near $y=0$ is then supplemented 
in Sec.~\ref{sec3} by a study of the
numerical diagonalisation of the transfer matrix for small widths but
for all values of $0<y<y_s$. This confirms the analytic results but also
shows how the eigenstates cross in a complicated manner as $y$ is
decreased. In particular, the low-lying states occur in different
sectors for $y\ll y_s$ and $y_s-y\ll y_s$. However, as $L$ is increased
all these crossings move towards $y=y_s$, indicating that in the limit
the scaling theory at all points $y<y_s$ is the same as at $y=0$, and
different from that at the special transition $y=y_s$. These numerical
results are limited by the small values of $L$ we use, and the fact
that, at $y=0$, a layer of sites is effectively removed at each
boundary. 

\section{Perturbation theory about the teflon point}\label{sec2}
We first consider the spectrum of the transfer matrix at the point $y=0$
and how it differs from that at the integrable ordinary point $y=x_c$.
The transfer matrix is constructed from the ensemble of \em oriented \em
self-avoiding loops. The number of up arrows minus the number of down
arrows crossing any given `time' slice is therefore conserved. This total
`charge' $Q$ therefore commutes with the transfer matrix, and the
eigenstates of the latter may be organised into sectors of a given $Q$.
All the way along the ordinary line $OS$ and at the special point $S$,
the largest eigenvalue $\Lambda_0$
of the transfer matrix is nondegenerate and 
lies in the $Q=0$ sector,\footnote{
Actually at the point $S$ the largest eigenvalues in the $Q=0$ and $Q=1$ sectors are
equal when $n=1$, with $x_1^{(s)} < 0$ for $n < 1$. Thus for $n < 1$
there is a point on the line $OS$ (below point $I$) where the
two eigenvalues cross. The central charge is still measured from the
largest eigenvalue in the  $Q=0$ sector.} 
and the corresponding
quantity $E_0\equiv-\ln\Lambda_0$ has the finite-size scaling form \cite{BCN86,A86}
\be
E_0=f_bL+2f_s-{\pi \zeta c(n)\over24L}+o(L^{-1}).
\ee
Here $\zeta = 2/\sqrt3$ is a lattice dependent scale factor, inserted
since $L$ counts the number of sites, rather than the linear dimension,
across the strip.
The bulk and surface free energies $f_b$ and $f_s$ are also
nonuniversal. In particular $f_s$ depends on the surface fugacity $y$.
However the central charge $c$ is universal, with \cite{BS93,BY95a,YB95b} 
\be
c(n)=1 - 6(g-1)^2/g,         \label{c}
\ee
where $n=-2\cos\pi g$. Similarly the excited
state eigenvalues $E_m\equiv-\ln\Lambda_m$ scale according to \cite{C84b}
\be
E_m-E_0={\pi \zeta x_m\over L}+o(L^{-1}),
\ee
where $x_m$ is one of the scaling dimensions of the allowed surface
operators. The terms in the $E_m$ which scale like $1/L$ may therefore
be referred to as the universal, scaling, part of the spectrum. 
There is always an
excited state in the $Q=0$ sector corresponding to the energy
operator of the $O(n)$ model. In the spin language this is the operator
$\no{S^*_aS_a}$, and it measures the local density of monomers, that
is, the probability that a given site is occupied. Along the ordinary
line $OS$, it has been argued that this operator has surface scaling dimension
$x_e=2$, and is in fact not an independent primary operator but is
proportional to the component $T_{\parallel\perp}$ of the stress tensor.
This has been verified by the exact solution at $y=x_c$. However, at the
special point $S$, this is not true, and it is found that 
$x_e^{(s)}=2/g-1$. 
There is another excited state in the $Q=0$ sector, degenerate with
the ground state in the $Q=2$ sector. The former corresponds to the
operator $\no{S^*_aS_b}$ (with $a\not=b$) while the latter
corresponds to $\no{S^*_aS^*_b}$ (or $\no{S_aS_b}$) which acts as a
source (or a sink) for a pair of oriented walks. These have the same
eigenvalue (and hence the same scaling dimension because they are
related by an $O(n)$ rotation (the full symmetry group of the problem.)
In fact, at $n=0$ these states are also degenerate with that
corresponding to the energy operator, as may be seen by a study of the
form of the transfer matrix.
In general the ground state in the charge $Q$ sector
corresponds to the operator $\no{S^*_{a_1}S^*_{a_2}\ldots S^*_{a_Q}}$,
which acts as a source for $Q$ oriented walks. 
It has scaling dimension \cite{BS93,YB95b} 
\be
x_Q=\ffrac{1}{4} g\, Q^2 + \ffrac{1}{2}(g-1) Q  \label{xord} 
\ee 
at the ordinary transition and \cite{BY95a,YB95b}
\be
x_Q^{(s)}= \ffrac{1}{4} g (Q+1)^2 - \ffrac{3}{2}(Q+1) +
\frac{9-(g-1)^2}{4 g}
\ee
at the special point. In general, these states will be
degenerate with states in the sectors $Q-2,Q-4,\ldots$ corresponding to
reversing some of the arrows.

\begin{figure}
\centerline{
\epsfxsize=4.5in
\epsfbox{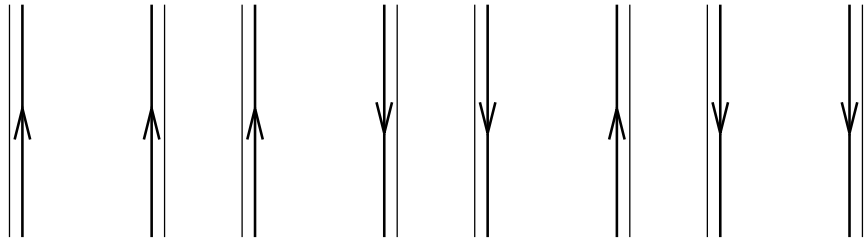}}
\caption{The ground state configurations at $y=0$. They occur in the
$|Q|=2$ and $Q=0$ sectors, the latter being doubly degenerate. In this
and subsequent diagrams, for clarity, the details of the honeycomb lattice 
are not shown. Thin lines represent the boundaries and thicker ones the
polymers. Also the sea of closed loops is not indicated.}
\label{gs}
\end{figure}

\begin{figure}
\centerline{
\epsfxsize=1.5in
\epsfbox{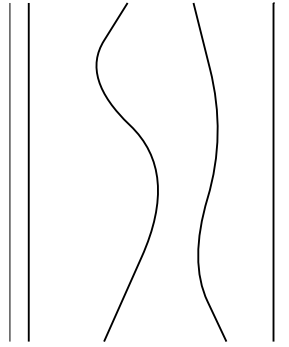}}
\caption{Configurations giving rise to the first excited state in the
$|Q|=2$ and $Q=0$ sectors, at $y=0$. Orientations are not shown:
depending on these, this state occurs in all even sectors with
$|Q|\leq4$.}
\label{1s}
\end{figure}

Now let us consider how this picture changes at the teflon point $y=0$.
Then all the boundary states must be occupied, and so there must
be a polymer lying along each boundary. However, these may have either
orientation. Once these are in place, the remaining problem is that of
a strip of width $L-2$ at the integrable ordinary point. Apart from
the boundary polymers, therefore, the spectrum is exactly that described
above. However, when these are included we now see that the ground
states in the $Q=0$ and $Q=\pm2$ sectors are identical. Moreover, that
in the $Q=0$ sector is doubly degenerate. This is
illustrated in Fig.~\ref{gs}.
The first excited state, with
finite-size scaling behaviour controlled by the scaling dimension
$x_e=2$, now occurs in all the even sectors with $|Q|\leq4$, with
different degeneracies, as shown in Fig.~\ref{1s}.

\begin{figure}
\centerline{
\epsfxsize=1.5in
\epsfbox{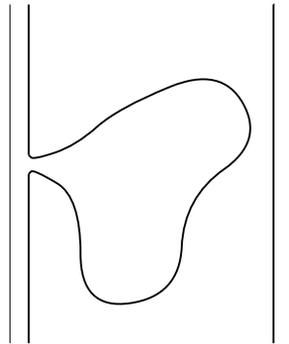}}
\caption{First order corrections to the ground state. Orientations are
not shown, but this contribution is diagonal and therefore independent
of the sector in which it occurs.}
\label{gsOy}
\end{figure}

\begin{figure}
\centerline{
\epsfxsize=3.0in
\epsfbox{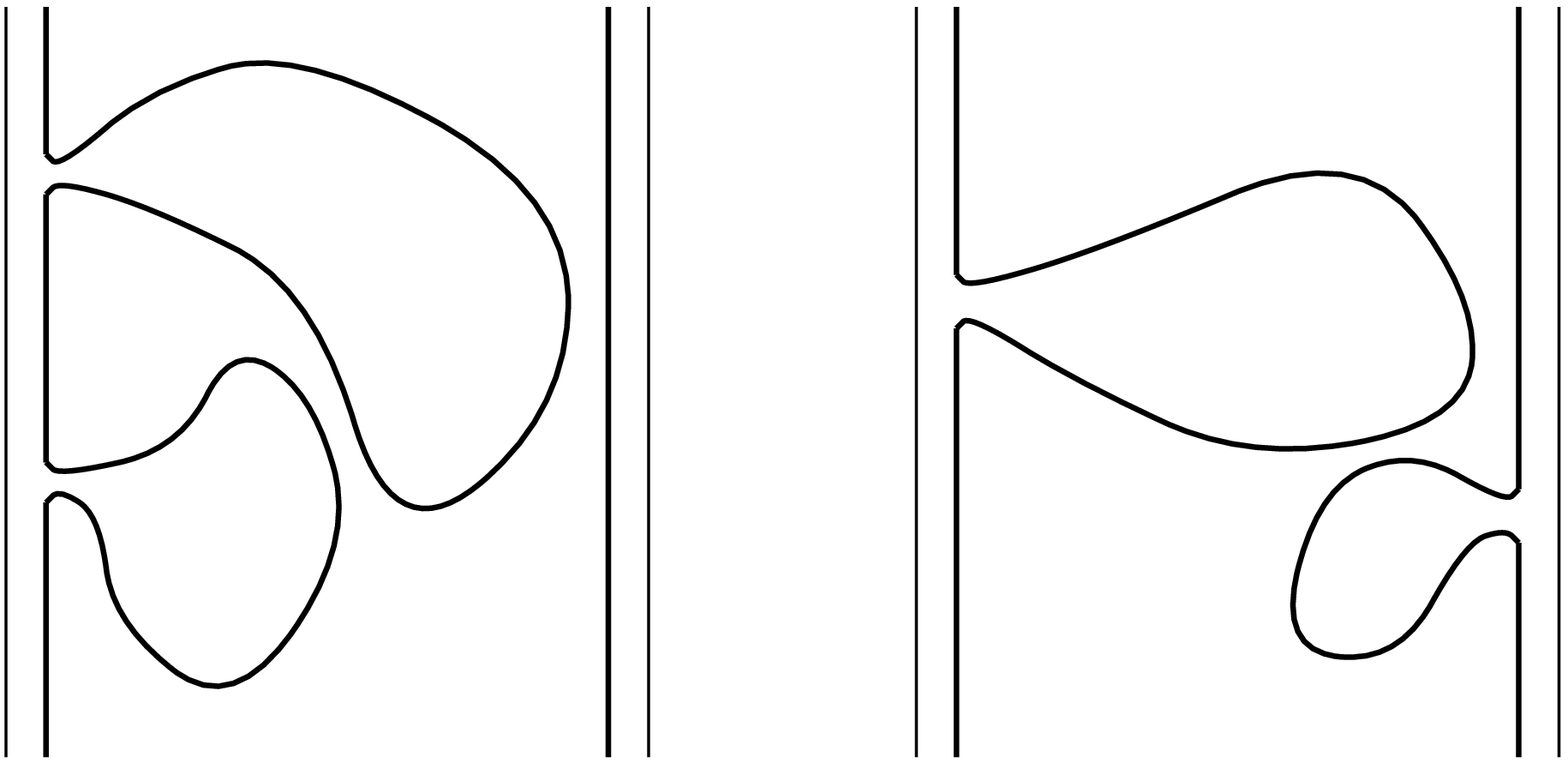}}
\caption{Second order contributions to the ground state. These are
diagonal in arrow space and therefore give equal shifts in all sectors.}
\label{gsOy2a}
\end{figure}

\begin{figure}
\centerline{
\epsfxsize=4.5in
\epsfbox{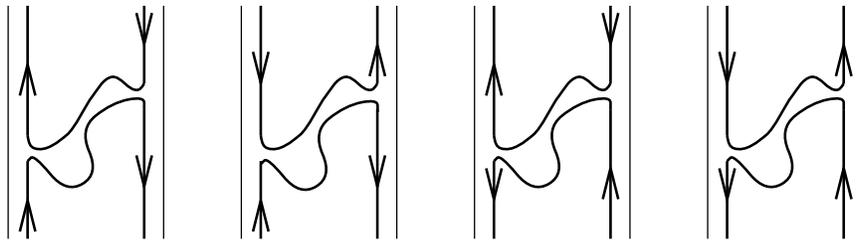}}
\caption{Second order contributions to the ground state which are
responsible for mixing in the $Q=0$ sector.}
\label{gsOy2b}
\end{figure}

When $y\not=0$, the first effect to $O(y)$ is to allow a single boundary
site to be vacant. This immediately allows one of the boundary polymers
to wander away from the surface at this point, as illustrated in
Fig.~\ref{gsOy}. 
Since the bulk is critical, it may in fact wander all
the way across to the other side and therefore modify the finite-size
scaling spectrum. This may be quantified by observing that if we now
ignore the two boundary layers, the effect is of a long polymer loop
attached at one point to the boundary. Such configurations are generated
by inserting the energy operator $S^*_aS_a$ near the boundary. Since
this has scaling dimension $x_e=2$, the $O(y)$ change in the free energy per
unit length will therefore scale like $L^{-2}$. The universal scaling
part of the spectrum is therefore unaffected, the change in $E_0$ being
$O(y/L^2)$. This also indicates that from the \rg\ point of view $y$ is
irrelevant, with eigenvalue $-1$. Thus flows from the vicinity of $y=0$
go
into the teflon fixed point, and all universal quantities like the
scaling part of the eigenvalue spectrum should be identical.
Note also that the $O(y)$ corrections
will be identical in the $|Q|=2$ and $Q=0$ sectors, so they remain
exactly degenerate to this order. The higher order corrections are more
interesting, however. They correspond to the configurations shown in
Figs.~(\ref{gsOy2a},\ref{gsOy2b}). 
Those in Fig.~\ref{gsOy2a} are
diagonal in the basis of unperturbed ground states, and therefore
produce shifts which, since the matrix elements are
independent of the orientation, are equal in all three sectors. However
some of the diagrams in Fig.~\ref{gsOy2b} produce mixing between the two
$Q=0$ states. Because of the extra $e^{\pm i\chi}$ factors introduced in
some of these configurations, the mixing matrix is proportional to 
$$
\pmatrix{e^{6i\chi}&1\cr
         1&e^{-6i\chi}\cr}
$$
with eigenvalues $0$ and $2\cos6\chi=n$. The two states in the $Q=0$
sector are therefore split at $O(y^2)$, but one remains degenerate
with those in the $|Q|=2$ sectors. At $n=0$ this splitting should
vanish. The order of magnitude of these second order shifts is expected
on the basis of the \rg\, to be $O(L^{-3})$. However, if they are
computed explicitly in terms of integrals over two-point functions in the
unperturbed theory, those corresponding to the first diagram in
Fig.~\ref{gsOy2a} diverge at short distances and must be
regulated by a cut-off of the order of the lattice spacing, giving rise
to contributions $O(y^2/L^2)$. Such non-universal
corrections to
finite-size scaling from higher order effects of an irrelevant operator
are standard. However, they do not affect the splitting coming from
the diagrams in Fig.~\ref{gsOy2b}, where the two insertions are on
opposite sides of the strip.

\begin{figure}
\centerline{
\epsfxsize=4.5in
\epsfbox{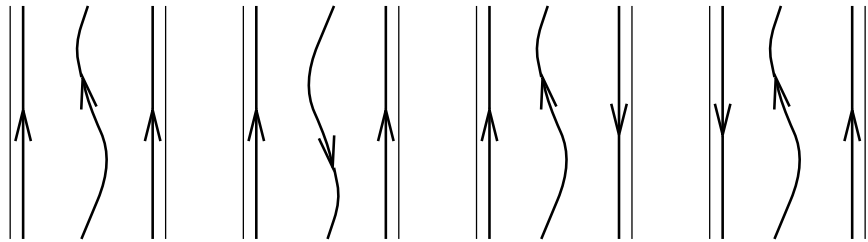}}
\caption{Lowest states in the sectors with $Q$ odd at $y=0$. They are all
degenarate: the first lies in the $Q=3$ sector, and the last three in
the $Q=1$ sector. Not shown are the charge conjugate configurations in
the $Q=-1,-3$ sectors.}
\label{Qodd}
\end{figure}

\begin{figure}
\centerline{
\epsfxsize=4.5in
\epsfbox{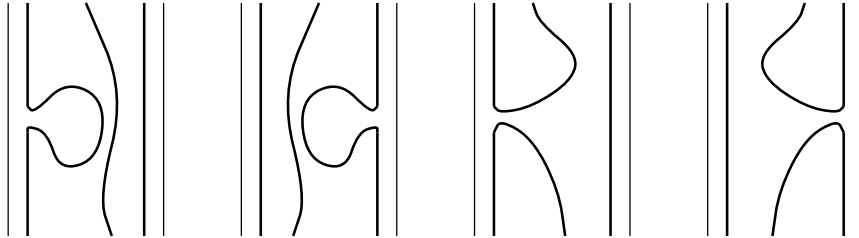}}
\caption{First order corrections to the lowest states in the odd $Q$
sectors. The first pair are diagonal, but the second pair are
responsible for the splitting of the $|Q|=1$ states.}
\label{QoddOy}
\end{figure}

Next we study the excited states in perturbation theory in $y$. These
effects are easiest to see in the odd $Q$ sectors. At $y=0$ the lowest
states in the $Q=\pm1,\pm3$ sectors are all degenerate, corresponding to
the configurations shown in Fig.~\ref{Qodd}. 
To $O(y)$, there are two
distinct types of configurations which enter, shown in
Fig.~\ref{QoddOy}. 
The first two are diagonal and therefore shift all
states equally, by an amount $O(y/L^2)$. The second pair induce mixing in
between the 3 degenerate states in the $|Q|=1$ sectors, however. As
above, the various $e^{\pm i\chi}$ factors must be taken into account,
and when this is done the mixing matrix in the $Q=1$ sector, for
example, is proportional to
$$
\pmatrix{e^{6i\chi}+e^{-6i\chi}&1&1\cr
         1&e^{6i\chi}&0\cr
         1&0&e^{-6i\chi}\cr}
$$
with eigenvalues $(n-1,0,n+1)$. Therefore, to $O(y/L^2)$, we expect to
find a single state in the $Q=1$ sector remaining degenerate with that in the
$|Q|=3$ sectors, while the remaining two split away. At $n=0$ this
splitting is equidistant. 

\section{Numerical evidence}\label{sec3}

In order to supplement the conformal perturbation theory results we have
numerically diagonalised the double-row transfer matrix of the vertex model.
The detailed investigation of degeneracies and splitting of the eigenvalues 
as a function of the boundary weight $y$ necessitates calculating many 
eigenvalues of the transfer matrix.
We have thus restricted ourselves to calculating the complete eigenspectrum for
small widths of up to $L=8$. This is sufficient to confirm both the order in $y$
and the finite-size dependence of a given splitting. Our numerical results are 
suggestive that the corrections to the finite-size scaling spectrum away from the 
teflon point are $O(y/L^2)$ and thus do not effect the scaling dimensions.

Consider first the teflon point, with $y=0$ on each side of the strip.
We confirm for finite $L$ that, apart from the shuffling of eigenvalues, 
the eigenspectrum is
equivalent to that on a strip of width $L-2$ with $y=x_c$ at the boundaries, 
i.e. at the integrable ordinary point.
The eigenvalues at $y=x_c$ appear with a higher degeneracy at $y=0$ 
(there are $3^L$ eigenvalues at $y=0$ compared to $3^{L-2}$ at $y=x_c$). 
The known exact results at the ordinary point can thus be used to infer exact
results at the teflon point. 
In contrast with the points $I$ and $S$, the 
largest eigenvalue in the $Q=0$ sector is seen to be 2-fold degenerate.
It is also degenerate with the largest eigenvalue in the $Q=2$ sector.
At $n=0$ the largest eigenvalue is given by 
$\Lambda_0 = (2 + \sqrt 2)^{L-2}$ and thus $c=0$.
For general $n$ the central charge is given by (\ref{c}), as calculated 
for the integrable ordinary point.
The leading thermal excitation in the $Q=0$ sector at $y=0$ corresponds to the
thermal excitation at the integrable ordinary point, and thus $x_e = 2$, 
independent of $n$.
More generally the scaling dimensions (\ref{xord}) also appear at $y=0$.
In particular, the magnetic scaling dimension is given by $x_h = x_1$. 

\begin{figure}
\centerline{
\epsfxsize=4.0in
\epsfbox{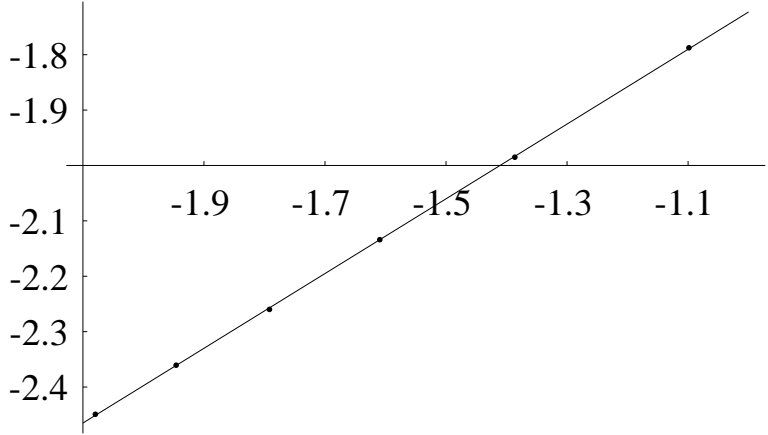}}
\caption{Log-log plot of $y_s - y_{cr}$ vs $1/L$ obtained from successive
estimates of the crossing point $y_{cr}(L)$ at $n=0$ for $L=3$ to $8$. 
The indicated line of best fit gives the cross-over 
exponent $y' \simeq 0.67$.} 
\label{loglog}
\end{figure}

Consider the largest eigenvalue in the $Q=0$ sector along the line $SO$ 
($y \ge y_s$) at $n=0$. At $n=0$, this corresponds to the state with no
arrows, which, for this value of $n$, is an exact eigenstate. Its
eigenvalue is given
exactly by $\Lambda_0 = (2 + \sqrt 2)^L a^2$, where $a=y/\sqrt{2 + \sqrt 2}$ 
is the surface free energy contribution. There are no further correction terms
in either $y$ or $L$. Clearly for finite $L$, $\Lambda_0 \to 0$ as $y \to 0$. 
Thus, for finite $L$, there is a point at which this eigenvalue crosses the continuation 
of the 2-fold degenerate ground state for $0 < y < y_s$.
For increasing $L$, the successive crossing points $y_{cr}(L)$ are 
expected to approach $y_s$ at a rate determined by the crossover exponent
$y'$ at the special point:
\be
y_s - y_{cr} \sim L^{-y'},
\label{cross}
\ee
where $y'=1-x_e^{(s)}=\frac23$ at $n=0$. 
A log-log plot of the above estimates, exhibited in Fig.~\ref{loglog}
reveals the value 
$y' \simeq 0.67$, in excellent agreement.

The 2-fold degeneracy is seen to be broken away from $n=0$, with one of the 
eigenvalues remaining degenerate with the largest eigenvalue 
in the $Q=2$ sector. According to the conformal perturbation theory, 
there should be a uniform shift $O(y/L^2)$ while the
splitting should be $O(y^2/L^2)$. 
Examination of these eigenvalues for small $y$ indicates that the 
$O(y)$ term tends to a constant as $L$ increases, being a surface 
energy contribution in $y$. For fixed $y$ the 
finite-size convergence to this value is consistent
with $O(1/L^2)$, however the available data is insufficient
to unambiguously determine the exponent.  
Nevertheless this term cancels in the finite-size estimates of the 
scaling dimensions.
The more important term for the scaling spectra is the $O(y^2)$ term 
governing the splitting of the degeneracy. 
The splitting is absent at $n=0$ and clearly seen to be $O(y^2)$ 
away from $n=0$. 
For finite $y$, the splitting term is seen to vanish at least 
as fast as $O(1/L^2)$.

The splitting of the 3-fold degenerate largest eigenvalue in the
$Q=1$ sector is clearly seen to be $O(y)$. Moreover, the splitting
is precisely proportional to the eigenvalues $(n-1,0,n+1)$ derived from
the mixing matrix. Denote the three split eigenvalues by
$\Lambda_1^{(1)} < \Lambda_1^{(2)} < \Lambda_1^{(3)}$ and define their
differences by $a=\Lambda_1^{(1)}-\Lambda_1^{(2)}$ and 
$b=\Lambda_1^{(2)} - \Lambda_1^{(3)}$. We observe the equal
splitting at $n=0$ with $b=0$ at $n=1$. Finite-size estimates of $a/b$ 
for the particular value $n=0.445\ldots$ ($g=\ffrac{10}{7}$)
are shown in Table \ref{tab}. They are seen to be in
good agreement with the expected value   
$a/b = - (n+1)/(n-1) = 2.60388$.
The middle eigenvalue $\Lambda_1^{(2)}$ remains degenerate with the largest
eigenvalue in the $Q=3$ sector.

\begin{table}[t]
\caption{Finite-size estimates of the splitting ratio $a/b$ for the
largest eigenvalue in the $Q=1$ sector. The expected value derived
from the mixing matrix is $a/b = 2.60388$.\label{tab}} 
\vspace{0.4cm}
\begin{center}
\begin{tabular}{||c|c|c|c||}
\hline
$L$  &$y=0.0001$ & $y=0.001$ & $y=0.01$    \\ \hline
  3  &2.60369    & 2.60202   & 2.58003   \\
  4  &2.60360    & 2.60107   & 2.57390   \\
  5  &2.60362    & 2.60133   & 2.57734   \\
  6  &2.60365    & 2.60163   & 2.58059   \\
  7  &2.60368    & 2.60187   & 2.58320   \\
  8  &2.60370    & 2.60207   & 2.58529   \\ \hline
\end{tabular}
\end{center}
\end{table}

\begin{figure}
\centerline{
\epsfxsize=4.0in
\epsfbox{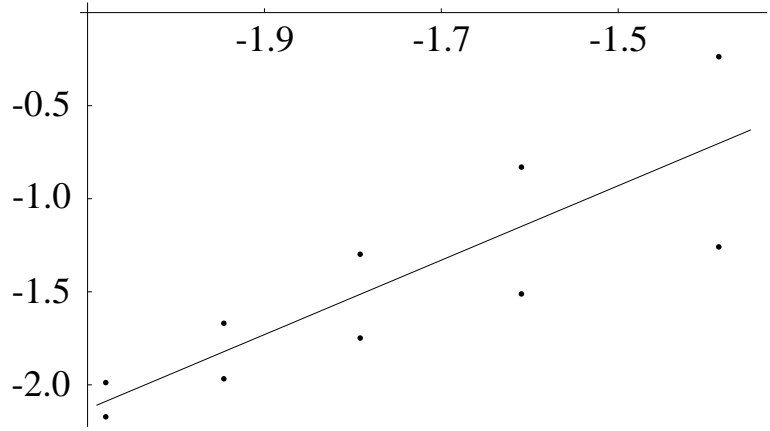}}
\caption{Log-log plot of the amplitudes $A$ (lower points) and $B$ 
(upper points) appearing in Eq.~(9) 
vs $1/L$ for $L=4$ to $8$. The solid line indicates
the expected $1/L^2$ convergence.} 
\label{amp}
\end{figure}

At fixed $y$ the amplitude of the $O(y)$ splitting is seen to decrease
with increasing $L$.
Consider, for example, $n=0$ where the splitting is equidistant, with 
\be
E_1(y) - E_0(y) \,=\, E_1(y=0) - E_0(y=0) + A \, y 
+ \left\{ \begin{array}{c} 
              B \\ 0 \\ -B 
          \end{array}   \right\} y + o(y^{-1}).    \label{AB}
\ee
The amplitudes $A$ and $B$ are constant for given $L$. However, 
they vanish as $O(1/L^2)$,
as can be seen in the log-log plot of Fig.~\ref{amp}.   
 
\begin{figure}
\centerline{
\epsfxsize=4.5in
\epsfbox{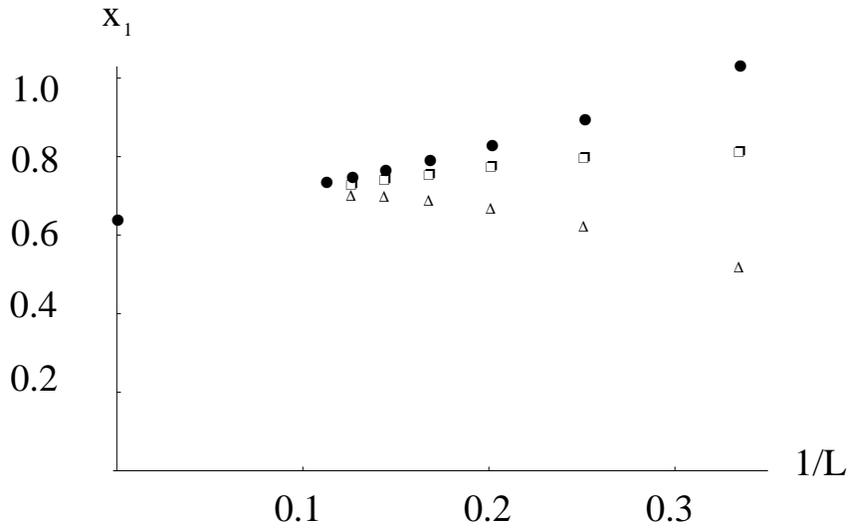}}
\caption{Estimates of the magnetic scaling dimension $x_1$ as a 
function of $1/L$ for $y=0$ ($\bullet$), $y=0.1$ ($\Box$) and 
$y=0.2$ ($\triangle$). For
each case the expected asymptotic value is $x_1=0.625$.}
\label{xest}
\end{figure}

The corresponding finite-size estimates of the magnetic scaling 
dimension $x_1$ are shown as a function of $1/L$ in Fig.~\ref{xest} 
for increasing values of $y$. In each case the observed trend
is again consistent with $O(y/L^2)$ convergence in the eigenvalues, 
i.e. the scaling dimensions being independent of $y$. 
We have observed similar behaviour in the estimates
of the thermal dimension $x_e=2$.  

\section{Conclusion}

We have studied the extraordinary transition in the 
two-dimensional $O(n)$ model for $n<1$ by
conformal perturbation theory and by finite-size
scaling of the transfer matrix eigenspectrum of a simple model
on the honeycomb lattice. We argued that the universal properties along
the line $0<y<y_s$ in the phase diagram shown in Fig.~\ref{pd}
are independent of $y$, being those at the teflon point $y=0$.
The surface scaling dimensions at $y=0$ are equal to those for the 
ordinary transition. There are, however, additional eigenspectra
degeneracies at $y=0$ compared to those along the line of the 
ordinary transition. 
The result for the magnetic scaling dimension at 
the extraordinary transition differs from the value $x_h=x_e=2$ 
obtained previously with fixed boundary conditions \cite{BEG89,BE94}. 
As we have explained, this is because the $O(n)$ model does not
order in the absence of a symmetry-breaking field when $n<1$.

Although the extraordinary transition we have discussed in this paper is
not relevant to the statistics of a single self-avoiding walk, or linear 
polymer, of fixed finite length, it may be realised within the
fixed fugacity ensemble. In this case a \em single \em polymer will simply
bind to the wall and not exhibit any interesting scaling behaviour. The
exponents we have discussed in the body of this paper, and which are observed
in the finite-size scaling spectrum, refer to the behaviour of a \em second \em
polymer in the vicinity of the surface. Thus our results are relevant to the
semi-dilute regime rather than the statistics of a single polymer.

It is also possible to consider mixed boundary
conditions, when one side of the strip has $0<y<y_s$ and the
other is either at the ordinary or the special point.
In that case, at the teflon point, the finite-size spectrum will be that
of a strip of width $L-1$ with either ordinary exponents or mixed 
ordinary-special exponents \cite{BY95b,YB95b,BO96}. The degeneracy of these
states will be different, however, and may be discussed in the manner as
in Sec.~\ref{sec2}, as may be the splittings which should occur away from
$y=0$.

\vskip 1cm\noindent
The authors thank A.~Owczarek for useful comments.
This work was begun while JC was a visitor at ANU under the 
Mathematical Sciences Research Visitors Program. It was continued
while MB was a visitor at Oxford under the Australian Academy of Science/
Royal Society Exchange Scheme, and completed while JC was a visitor at
the Institute for Theoretical Physics, Santa Barbara.
The work of MB has also been supported by the Australian Research Council,
and of JC by the EPSRC through Grant GR/J78327, and the NSF through Grant
PHY94-07194.



\end{document}